\documentstyle[epsfig]{mn}
\def\be{\begin{equation}} 
\def\ee{\end{equation}}

\def\HI{\hbox{H~$\scriptstyle\rm I\ $}}

\def\nhi{{N_{\rm HI}}}  
 
\def\gsim{\lower.5ex\hbox{\gtsima}} 
\def\lsim{\lower.5ex\hbox{\ltsima}} \def\gtsima{$\; \buildrel > \over 
\sim \;$} \def\ltsima{$\; \buildrel < \over \sim \;$} \def\prosima{$\; 
\buildrel \propto \over \sim \;$} \def\gsim{\lower.5ex\hbox{\gtsima}} 
\def\lsim{\lower.5ex\hbox{\ltsima}} 
\def\simgt{\lower.5ex\hbox{\gtsima}} 
\def\simlt{\lower.5ex\hbox{\ltsima}} 
\def\simpr{\lower.5ex\hbox{\prosima}}   
  
\def\gtsima{$\; \buildrel > \over \sim \;$} 
\def\ltsima{$\; \buildrel < \over \sim \;$} 
\def\gsim{\lower.5ex\hbox{\gtsima}} 
\def\lsim{\lower.5ex\hbox{\ltsima}} 
\def\simgt{\lower.5ex\hbox{\gtsima}} 
\def\simlt{\lower.5ex\hbox{\ltsima}} 
\def\simpr{\lower.5ex\hbox{\prosima}}

\def\Zcr{Z_{\rm cr}}

\def\Msun{M_{\odot}} %{\,{\rm M_\odot}}
\def\Zsun{Z_{\odot}}

\def\E3{{\cal E}_{\rm g}^{III}}

\title{First stars in Damped Lyman Alpha systems}
\author[S. Salvadori \& A. Ferrara]{Stefania Salvadori$^{1}$\thanks{E-mail:salvadori@astro.rug.nl} \& Andrea Ferrara$^{2}$\\ 
$^{{1}}$ Kapteyn Astronomical Institute, Landleven 12, 9747 AD Groningen, The Netherlands\\ 
$^{2}$ Scuola Normale Superiore, Piazza dei Cavalieri 7, 56126 Pisa, Italy}

\begin{document} 
\date{} 
\pagerange{\pageref{firstpage}--\pageref{lastpage}} \pubyear{} 
\maketitle 

\label{firstpage} 
\begin{abstract}
%%%%%%%%%%%%%%%%%%%%%%%%%%%%%%%%%%%%%%%%%%%%%%%%%%%%%%%%%%%%%%%%%%%%%%%%%%%%%%
In order to characterize Damped Lyman Alpha systems (DLAs)  
potentially hosting first stars, we present a novel approach to investigate  
DLAs in the context of Milky Way (MW) formation, 
along with their connection with the most metal-poor stars and local dwarf
galaxies. The merger tree method previously 
developed is extended to include inhomogeneous reionization and metal mixing, 
and it is validated by matching both the Metallicity Distribution Function of 
Galactic halo stars and the Fe-Luminosity relation of dSph galaxies. 
The model explains the observed $\nhi$-Fe relation of DLAs along 
with the chemical abundances of [Fe/H]$<-2$ systems. In this picture, the recently
discovered $z_{abs}\approx 2.34$ C-enhanced DLA (Cooke et~al. 2011a), pertains to 
a new class of absorbers hosting first stars along with second-generation long-living 
low-mass stars. These ''PopIII DLAs'' are the descendants of H$_2$-cooling minihalos 
with $M_h\approx 10^7 \Msun$, that virialize at $z > 8$ in neutral, primordial 
regions of the MW environment and passively evolve after a short initial 
period of star formation. The gas in these systems is warm $T_g \approx (40-1000)$~K, 
and strongly C-enriched by long-living, extremely metal-poor stars of total mass 
$M_*\approx 10^{2-4}\Msun$. 
%%%%%%%%%%%%%%%%%%%%%%%%%%%%%%%%%%%%%%%%%%%%%%%%%%%%%%%%%%%%%%%%%%%%%%%%%%%%%%%%%%%%%%%
\end{abstract} 
%%%%%%%%%%%%%%%%%%%%%%%%%%%%%%%%%%%%%%%%%%%%%%%%%%%%%%%%%%%%%%%%%%%%%%%%%%%%
\section{Motivation}   
\label{intro}
%%%%%%%%%%%%%%%%%%%%%%%%%%%%%%%%%%%%%%%%%%%%%%%%%%%%%%%%%%%%%%%%%%%%%%%%%%%%
Damped Ly$\alpha$ absorption systems (DLAs) are high-column density 
neutral gas reservoirs, $\nhi \geq 10^{20.3}$cm$^{-2}$, observed at 
intermediate redshifts, $z\leq 5$, in the spectra of distant quasars. 
Although their nature is still unclear (Pettini 2004), the key role of 
DLAs to understand galaxy formation  (Wolfe, Gawiser \& Prochaska 2005) 
is widely recognized. So far, more than $1000$ DLAs have been observed 
and the iron abundance measured in $\approx 150$ systems 
$[Fe/H]\approx[-3.5,-0.5]$ (Prochaska et~al. 2007). 
Among these, very metal-poor (VMP) DLAs with [Fe/H]$<-2$, 
can be used to study the initial phases of heavy element enrichment 
of the interstellar medium (ISM) of early galaxies. Indeed, if VMP 
stars observed today in the Galactic halo and in nearby dwarf 
spheroidal galaxies (dSphs) are the living fossils of the first 
stellar generations, VMP DLAs may well constitute the gas reservoir 
out of which such pristine stellar populations formed. 
Following the medium resolution study of VMP DLAs by Penprase et~al. (2010), 
Cooke et~al. (2011b) have recently presented a high spectral resolution 
sample, including $22$ VMP systems. In these DLAs [C/O]$\approx -0.3$ and 
[Fe/O]$\approx -0.4$ independently of [Fe/H] and with little scatter, 
in agreement with measurements in VMP Galactic halo stars (Fabbian 
et~al. 2009). So far, the only exception to this general trend is 
represented by a DLA with [Fe/H]$ \approx -3$ and $\nhi = 10^{20.55 
\pm 0.10}$cm$^{-2}$ observed in the spectrum of the QSO J0035-0918 
at $z_{abs}\approx 2.34$ (Cooke et~al. 2011a). This system has 
[C/Fe]$=1.53$, i.e. $\approx 20$ times larger than any other DLA. 
Moreover, its abundance pattern shows a clear 'odd-even' effect and 
is consistent with the predictions for the yields of $Z=0$ faint 
supernovae (SN) with $m_*\approx 25\Msun$ (Kobayashi et al.~2011).  
Are we observing for the first time a DLA whose gas retains the imprint
left by the first stars? 

To characterize DLAs potentially hosting the first stars or their 
ashes, we propose a novel approach that simultaneously follows the 
evolution and chemical properties of DLAs and their connection with the 
most metal-poor stars and galaxies observed in the Local Universe based
on the results obtained using the merger-tree code GAMETE (Salvadori, 
Schneider \& Ferrara 2007, SSF07; Salvadori, Ferrara \& Schneider 2008; 
Salvadori, \& Ferrara 2009, SF09). 
%%%%%%%%%%%%%%%%%%%%%%%%%%%%%%%%%%%%%%%%%%%%%%%%%%%%%%%%%%%%%%%%%%%%%%%%%%%
\section{Model summary and validation}
%%%%%%%%%%%%%%%%%%%%%%%%%%%%%%%%%%%%%%%%%%%%%%%%%%%%%%%%%%%%%%%%%%%%%%%%%%%
\begin{figure*}      
\begin{centering}
  \includegraphics[height=.58\columnwidth]{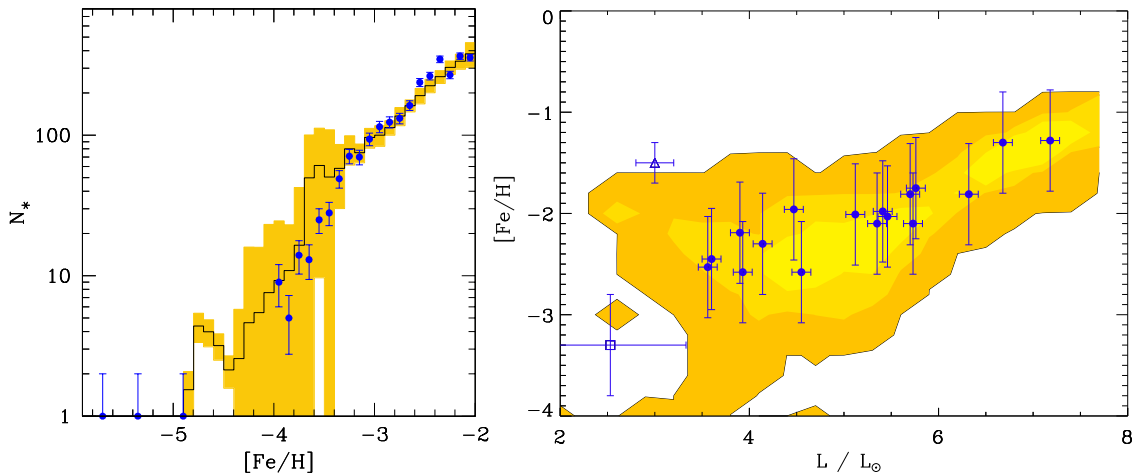}
  \includegraphics[height=.73\columnwidth]{FeL.epsi}
    \caption{{\it Left panel}: observed (points with error-bars) and
simulated (histogram) metallicity distribution function of Galactic
halo stars. The data points are the sample by Beers \& Christlieb (2006)
with the inclusion of the three hyper-iron poor stars (Christlieb et al.
2002, 2006; Frebel et al. 2005). The histogram is the average MDF value 
over 50 merger history of the MW, re-normalized to the total number of 
observed stars with [Fe/H]$\leq −2$.
The shaded area represents $\pm 1\sigma$ errors. {\it Right panel}: observed
(points with error-bars) and simulated (contours) iron-luminosity relation
for the MW dSph galaxies. The data points are by Kirby et~al. (2008) with
the inclusion of the two faintest dSphs (Willman et~al. 2005, Geha et~al. 
2009). The colored shaded areas correspond to regions that include the 
$(99,95,68)\%$ of the total number of possible dSph candidates selected 
in 50 merger histories of the MW.}
\end{centering}
\end{figure*}
%%%%%%%%%%%%%%%%%%%%%%%%%%%%%%%%%%%%%%%%%%%%%%%%%%%%%%%%%%%%%%%%%%%%%%%%%%%%%%%
The model key points can be summarized as follows (see SSF07 and SF09 for details):
\begin{enumerate}
\item Hierarchical merger histories of a MW-sized dark matter (DM) 
halo are reconstructed from $z=20$ via a Monte Carlo algorithm based 
on the Extended Press-Schechter theory (SSF07). 
\item Star formation (SF) is followed along the tree in halos exceeding
a mass threshold, $M_{sf}$, whose evolution (Fig.~2) is governed by:
(a) the photo-dissociating Lyman-Werner (LW) background,
quenching H$_2$-formation in $T_{vir} < 10^4$~K minihalos; 
(b) the gas temperature in ionized regions of the Galactic Medium (GM), 
preventing gas-infall in halos with $T_{vir}$ lower than a threshold 
value, $T_{th}$. We assume that  SF is active in $T_{vir} > T_{th} =
2\times 10^3$~K minihalos at $z > 10$ (Dijkstra et~al. 2004). At lower $z$, 
$T_{th}$ (and hence $M_{sf}$) is assumed to linearly increase up to the value 
set by the end of reionization $T_{th}\approx 2\times 10^4$~K (Kitayama et~al. 
2000) for $z<z_{rei}=6$.
\item {\it Inhomogeneous} reionization is modeled by random 
sampling the reionization history implied by $M_{sf}(z)$ to switch off (on) 
gas accretion in minihalos that form in ionized (neutral) regions. 
\item The SF rate is taken to be proportional to the mass of cold gas, 
$\dot {M}=\epsilon_* M_g/t_{ff}$, where $\epsilon_*$ is the SF efficiency 
and $t_{ff}$ the halo free-fall time. In minihalos $\epsilon_*$ is reduced 
as $\epsilon=\epsilon_*[1+(T_{vir}/2\times 10^4$K$)^{-3}]^{-1}$ due to the 
ineffective cooling by H$_2$ molecules (SF09). Low-mass, Population II 
(PopII) stars form according to a Larson IMF when the gas metallicity exceeds 
{\it $\Zcr = 10^{-3.8} \Zsun$} (Schneider et~al. 2002). At lower metallicity, 
PopIII stars form with a reference mass value $m_*=25\Msun$ and explosion 
energy $E_{SN}=10^{51}$~erg consistent with  {\it faint} SNe. \item The abundance 
evolution of different chemical  elements\footnote{For $m_* < 8 \Msun$ stars we 
use yields by van den Hoek \& Groenewegen (1997) ($Z \geq 10^{-3}$) and 
by Meynet \& Maeder (2002) ($Z \leq 10^{-5}$); for more massive stars we use Woosley 
\& Weaver (1995) with a systematic halving of the Fe yield (Timmes, Woosley \& Weaver 1995); 
yields for faint SNe are from Kobayashi et al. (2011).} (from C to Zn) 
is traced in both the ISM and in the GM by taking into account mass- and 
metallicity-dependent stellar evolutionary timescales (Raiteri, Villata \& 
Navarro 1996) and SN feedback (Salvadori, Ferrara \& Schneider 2008).
\item To account for the {\it incomplete mixing} of SN ejecta within the ISM of 
gas-poor galaxies, gas outflows have a metallicity 
$Z_w = Z_{ISM} + \eta(M_g) {M_Z}/{M_{ej}}$, where $M_{Z}$ is the mass of 
newly formed metals, $M_{ej}$ is the mass of gas ejected out of the halo, 
and $\eta$ is a function of the gas mass $\eta=0.5+0.65\tanh[({\rm log10}
(M_g)-7.0)/2.0]$ that varies in the range $\eta=[0,1]$. 
\item The probability for newly formed halos to reside in a GM metal enriched region 
is $P(z)=Q_Z/Q_{\delta>\delta_c}$, where $Q_Z(z)=1-{\rm exp}(\Sigma_i 4\pi R_b^3(i)/V_{MW}(z))$
is the filling factor of metal bubbles within the MW physical volume\footnote{We assume 
$V_{MW}(z)=5(1+z)^{-3}$~Mpc$^3$.}, and $Q_{\delta>\delta_c}(z)$ is the volume filling factor 
of fluctuations with overdensity above the critical threshold, $\delta > 
\delta_c=1.686$, for the linear collapse (Miralda-Escud\'e, Haehnelt \& 
Rees 1999). The latter quantity describes the abundance 
of high-density regions, in which metals {\it first} penetrate (Tornatore, Ferrara \& Schneider 2007).
Objects in enriched (primordial) regions are assigned an initial metallicity 
$Z_{vir}=Z_{GM}/Q_Z$ ($Z_{vir}=0$).
\end{enumerate}
%%%%%%%%%%%%%%%%%%%%%%%%%%%%%%%%%%%%%%%%%%%%%%%%%%%%%%%%%%%%%%%%%%%%%%%%%%%
The model is calibrated by best-fitting the SF and feedback efficiencies to 
reproduce the global properties of the MW (stellar/gas mass and metallicity). 
In Fig.~1 we show that the average metallicity distribution of [Fe/H]$<-2$ 
stars over 50 possible MW merger histories matches the observed Galactic halo 
MDF. In the same Figure, the theoretical Fe-Luminosity relation for dSph
galaxies has been obtained by evolving in isolation star-forming halos with 
$M_h<M_{2\sigma}$, candidates to remain MW satellites (Diemand, Madau \& Moore 
2001; SF09 for details). In Fig.~2 $M_{sf}(z)$ is compared with that 
obtained by using data-constrained reionization models.
To this end we compute the dissociating LW background intensity associated 
with an early/late reionization history (Gallerani, Choudhury \& Ferrara 2006) 
by using the results provided in Fig.~6 by Ahn et~al. (2009); we then convert 
the flux into a critical mass following Machacek, Brian \& Abel (2001) and 
extrapolate to higher masses. At each redshift 
our $M_{sf}$ is within the range allowed by the two reionization histories
(gray shaded area).           
%%%%%%%%%%%%%%%%%%%%%%%%%%%%%%%%%%%%%%%%%%%%%%%%%%%%%%%%%%%%%%%%%%%%%%%%%%%%%%%
\section{The dwarf galaxy zoo}
%%%%%%%%%%%%%%%%%%%%%%%%%%%%%%%%%%%%%%%%%%%%%%%%%%%%%%%%%%%%%%%%%%%%%%%%%%%%%%%
\begin{figure}
 \begin{centering}
   \includegraphics[width=0.49\textwidth]{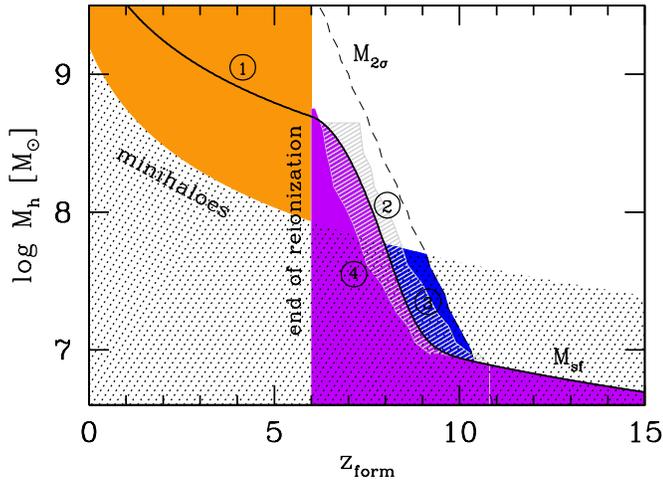}
  \caption{Dark matter content of MW satellites ($M_h < M_{2\sigma}$) 
as a function of their formation redshift. The two lines show the evolution 
of: the minimum mass of star-forming haloes, $M_{sf}$ (solid), the halo mass 
corresponding to $2\sigma$ density peaks (dashed). The dotted shaded area 
identifies the region populated by H$_2$-cooling minihalos with $T_{vir}<
10^4$~K. The number labels and colors specify different populations of 
satellites (see the text). The gray shaded area delimits the evolution 
of $M_{sf}$ predicted by assuming an early (upper limit) and a late 
(lower limit) reionization history (Gallerani et~al. 2007).} 
  \label{fig:zoo}
 \end{centering}
\end{figure}
%%%%%%%%%%%%%%%%%%%%%%%%%%%%%%%%%%%%%%%%%%%%%%%%%%%%%%%%%%%%%%%%%%%%%%%%%%%%%%
We now apply our model to DLAs starting from the origin of the C-enhanced, 
[Fe/H]$ \approx -3$ DLA at $z_{abs}=2.34$. Two possibilities must be considered:
(a) a recently virialized halo, later merging into the MW, or (b) a 
satellite galaxy which assembled at earlier epochs and evolved in isolation. 
The first possibility can be excluded on the basis of a metallicity argument: 
metal-free regions of the MW environment are expected to disappear at $\langle 
z\rangle \approx 8.7\pm  1.5$ when $P = Q_{Z}/Q_{\delta>1.686} > 1$. Afterward, 
newly virializing halos form from metal enriched GM gas, whose initial abundances 
[X/H]$_{vir}=\langle$[X/H]$\rangle_{GM}/Q_Z$. Since at $z=2.3$, $Q_Z = 0.989\pm 
0.005$ and $\langle$[Fe/H]$\rangle = -1.28\pm 0.02$, proto-galaxies with an initial 
iron abundance [Fe/H]$ < -1.3$ cannot form anymore.

In the second hypothesis, the dark matter (DM) halo hosting the DLA must be
associated to density fluctuations $<2\sigma$ at its final assembling epoch 
in order to evolve in isolation (no further merger or accretion) and become 
a satellite. By inspecting Fig.~2 we can distinguish among different 
populations of objects that satisfy this condition: 
(1) Ly$\alpha$-cooling halos that assembled {\it after} the end of 
reionization; Fornax and LeoI, the most luminous among the observed dSphs, 
are predicted to belong to this domain;
(2) star-forming Ly$\alpha$-cooling halos that assembled {\it before}
the end of reionization; classical dSphs, $L>10^5L_{\odot}$, such as
Sculptor (SF09) are members of this population.
(3) H$_2$-cooling, inefficiently star-forming minihalos that appeared at 
$z\approx 8-10$; this is the domain of ultra-faint dSphs (SF09).
(4) {\it "sterile''} halos that formed before the end of reionization 
and unable to trigger SF ($M_h<M_{sf}$). We selected the satellite candidates in 50 possible merger histories of 
the MW, and evolve them in isolation down to redshift $z=2.34$, 
when we compute their \HI column density,  $\nhi$, assuming 
that the gas is {\it fully} neutral with a molecular weight $\mu = 1.22$:
\begin{eqnarray}\nonumber
\nhi \approx 2n_{\rm HI}r_g \approx \frac{3} {2\pi}\;\frac{M_g/\mu m_p}{\alpha^2 r_{vir}^2} \newline
\end{eqnarray}
\begin{equation}\nonumber
= 5.2 \times10^{13}\;\frac{M_g}{\alpha^2}\;\Big(\frac{M_h}{10^8 M_{\odot}}\Big)^{-2/3}\;\Big(\frac{10}{1+z_{form}}\Big)^{-2}\;{\tt{cm}^{-2}}
\end{equation}
the gas radius is written as $r_g=\alpha r_{vir}$, with $\alpha=0.18$ 
for star-forming galaxies, that have presumably developed a disk 
(Ferrara, Pettini \& Shchekinov 2000), and $\alpha = 0.8$ otherwise. 
DLAs are canonically defined as systems with $\nhi > 2\times 10^{20}$
cm$^{-2}$ (Wolfe et al.~2005). Interestingly, we find that systems which 
populate Zone (2), i.e. the progenitors of classical dSphs, have already 
exhausted most of their gas by $z=2.34$, i.e. $\log (\nhi/{\tt cm}^{-2}) 
\ll 20.3$. On the other hand, many objects among those formed 
in Zones (1), (3) and (4), have \HI column densities compatible with DLAs. 
We will then focus on these candidates.
%%%%%%%%%%%%%%%%%%%%%%%%%%%%%%%%%%%%%%%%%%%%%%%%%%%%%%%%%%%%%%%%%%%%%%%%%%%%%
\begin{figure*}   
\begin{centering}
  \includegraphics[width=0.9\columnwidth]{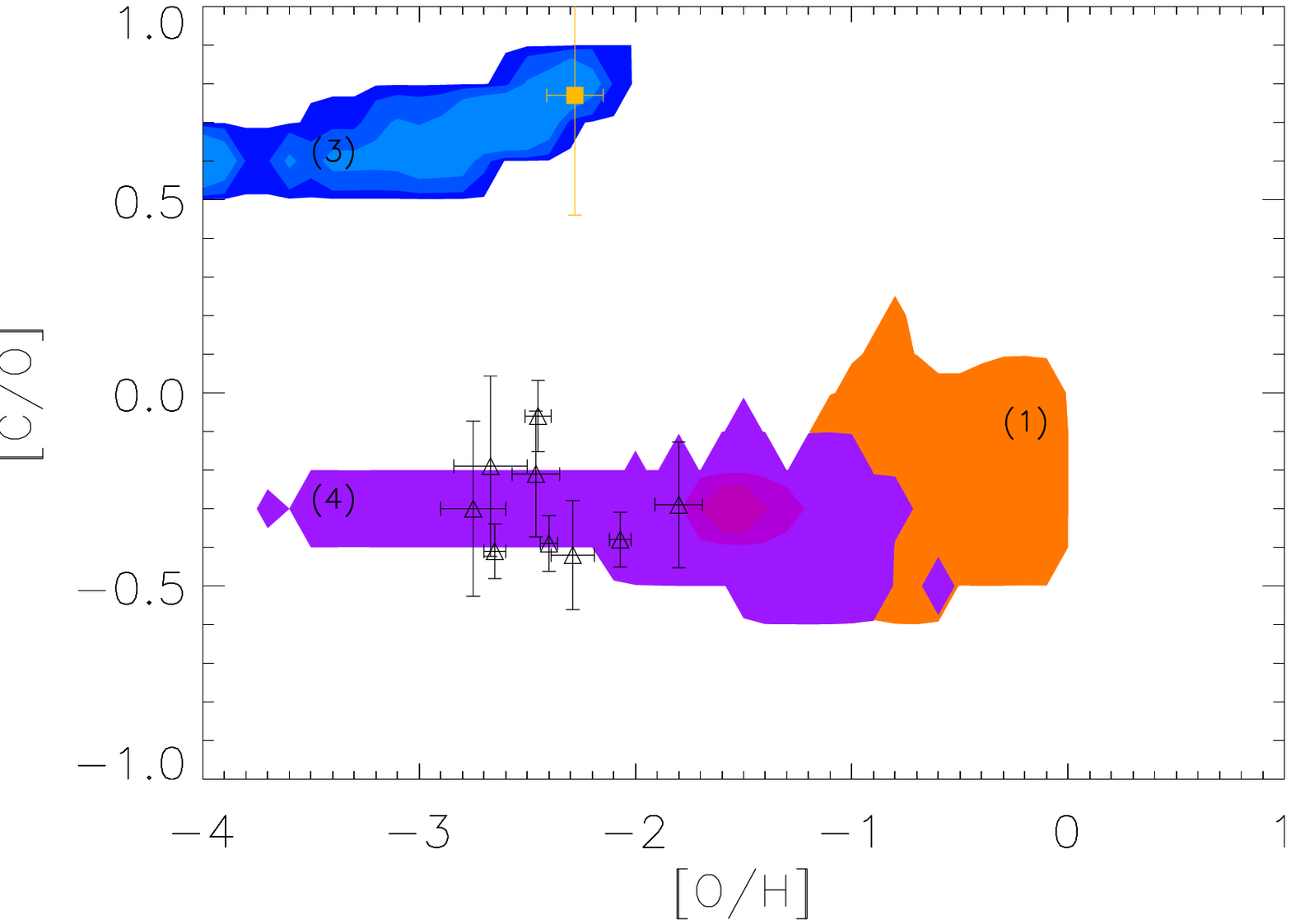}
  \includegraphics[width=0.9\columnwidth]{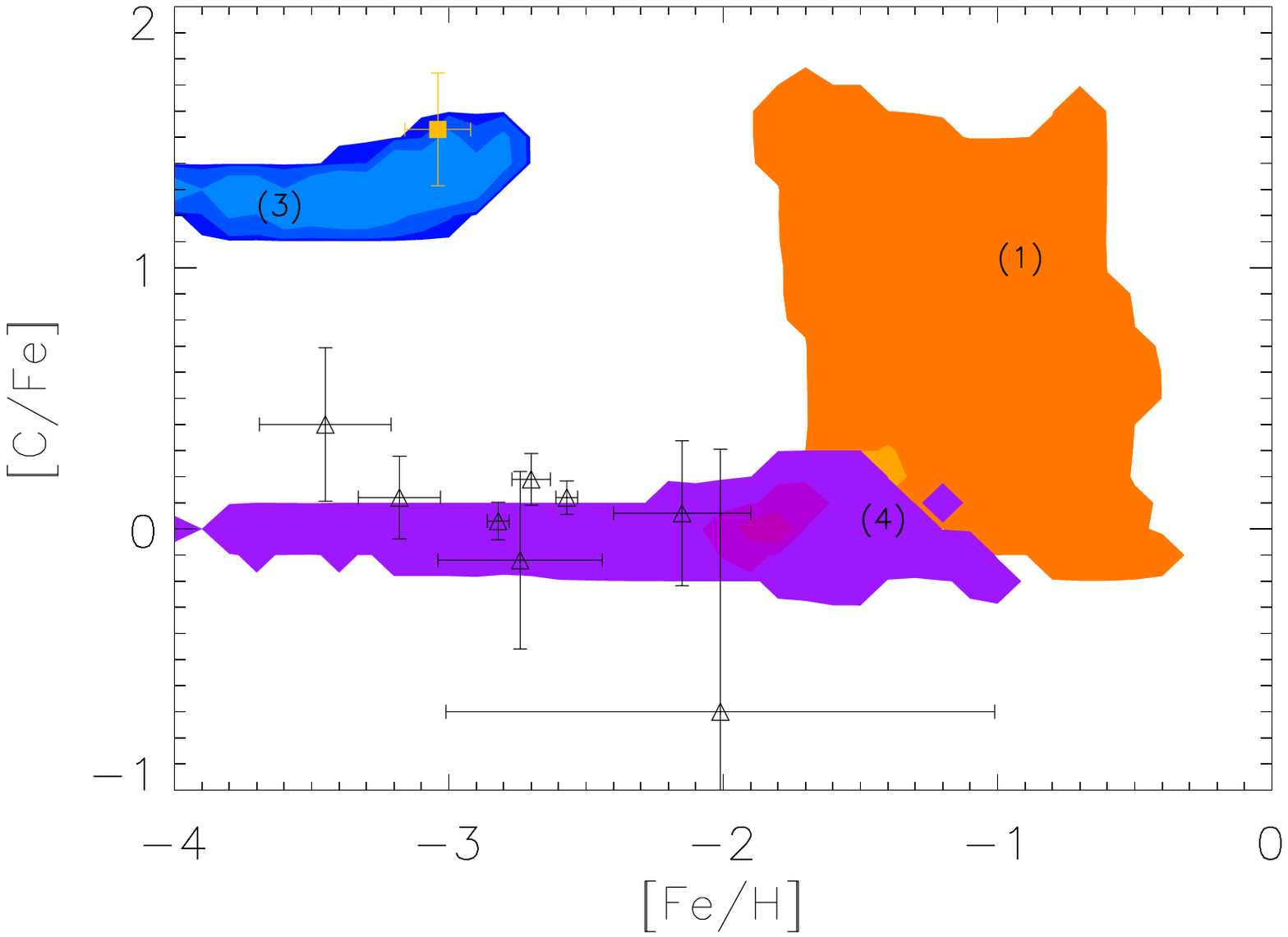}
    \caption{Gas abundances of DLAs with respect to the solar values 
  (Asplund et~al. 2009). Points are the observed values in VMP DLAs at 
  $z_{abs}=(2.07-4.21)$ (triangles, Cooke+2011b) and in the C-enhanced 
  DLA at $z_{abs}=2.34$ (square, Cooke+2011a). The color shaded areas show 
  the results of the model for the three different populations of DLAs 
  at $z\approx 2.34$: PopIII (blue), very metal-poor (violet), and 
  metal-poor DLAs (orange). For each population the intensity of the 
  colors correspond to regions containing the ($99,95,68$)\% of DLAs.}
\end{centering}
\end{figure*}
%%%%%%%%%%%%%%%%%%%%%%%%%%%%%%%%%%%%%%%%%%%%%%%%%%%%%%%%%%%%%%%%%%%%%%%%%%%%
\section{DLAs and first stars}
%%%%%%%%%%%%%%%%%%%%%%%%%%%%%%%%%%%%%%%%%%%%%%%%%%%%%%%%%%%%%%%%%%%%%%%%%%%%%
A comparison between the properties of candidate and observed DLAs at 
$z\approx 2.34$ is given in Fig.~3 and Fig.~4. Objects formed in different Zones result in distinct DLA populations, differing in their chemical abundances and $\nhi$. 
Systems in Zone (3) are segregated in the C-rich, 
metal-poor, low-$\nhi$ regions of the plots, matching the properties of the 
C-enhanced DLA. What is the origin of these systems?

According to our model, C-enhanced DLAs are hosted by $M_h\approx 
10^{6.9-7.2} \Msun$ H$_2$-cooling minihalos, which virialized before
the end of reionization out of {\it metal-free}, neutral regions of the MW 
environment. Because of their low DM content the star formation history of 
such a primordial proto-galaxies is extremely short. Some tens of Myrs after 
the onset of star-formation the intensity of LW background becomes high enough 
($M_{sf} > M_h$) to suppress further SF activity, thus turning them into ''sterile'' minihalos. Metal-free stars are hosted by this population of DLAs, to whom we will hereafter refer as PopIII DLAs. 

The metal-enrichment by first stars gently proceeds in these objects, 
mainly because of ineffective cooling by H$_2$ molecules and their 
low gas-mass content, $M_g\approx 10^{5.5}-10^{6.3}\Msun$, favoring 
metals and gas loss ($M^{ej}_g \approx 10^{5-6}\Msun$). As soon as 
$Z > \Zcr = 10^{-3.8} \Zsun$, however, ''normal'' PopII stars can 
form and contribute to enrichment. Given the yields by faint SNe, 
the PopIII-to-PopII transition occurs when [Fe/H]$_{cr} \approx -4.8$, 
implying that most of the iron observed in these systems 
originates from these second generations of stars. The mass of relic stars 
in PopIII DLAs is expected to vary between $M_*\approx 10^{2-4} \Msun$, 
the most star-rich systems being the most enriched ones. The rise of the 
C abundance at increasing metallicity (Fig.~3) reflects the gradual 
contribution by low-metallicity AGB stars, mainly producing C, some O, 
and limited amounts of N (for systems with [Fe/H]$=-3.0\pm 0.2$ we find 
[C/H]$=-1.6\pm 0.5$, [N/H]$=-3.8\pm 0.9$ and [O/H]$=-2.4\pm 0.4$).
The associated $\nhi$ decrease (Fig.~4) is caused by gas consumption 
in the most star-rich systems, due to both astration and gas loss. 
Note that our model does not predict any [C/O] increase towards low 
[O/H] values, as reported by medium resolution observations (Penprase 
et~al. 2010).

The properties of VMP DLAs are well reproduced (Fig. 3) by [Fe/H]$ < -2$ 
{\it starless} systems formed in Zone (4), {\it passively evolving} since 
their assembly epoch, $6 < z < 10$. They form through merging of primordial and 
metal-enriched progenitors, virialized out of neutral regions of the MW 
environment. The metallicity spread of these systems depends on the gas 
enrichment in progenitor minihalos. Abundance ratios, instead, closely 
reflect those of the GM at the time of formation. Although the overall 
GM metallicity increases with time, abundances ratios get locked to the 
dominant stellar population which contribute to the enrichment, i.e. 
{\it type II SNe}. As a consequence, [C/O] and [C/Fe] ratios show
little dispersion and are independent of $Z$, but also of redshifts,
because of the passive evolution of these starless DLAs. These findings
are in perfect agreement with the new observational results by Becker et~al.
(2011).
%%%%%%%%%%%%%%%%%%%%%%%%%%%%%%%%%%%%%%%%%%%%%%%%%%%%%%%%%%%%%%%%%%%%%%%%%%%%%%%
\begin{figure}
\begin{centering}
  \includegraphics[width=0.99\columnwidth]{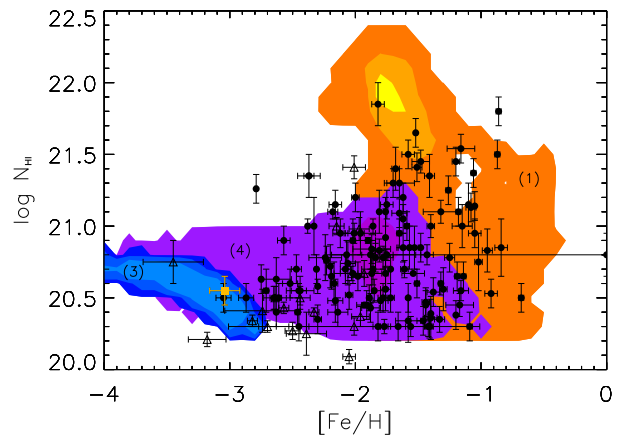}
    \caption{$\nhi$ distribution vs. iron abundances of DLAs. 
Points with error-bars are observed values for metal-poor DLAs at 
$z_{abs}=(1.9-4.9)$ (circles, Prochaska et~al. 2007), VMP DLAs at 
$z_{abs}=(2.07-4.21)$ (triangles, Cooke+2011b) and for the C-enhanced DLA
at $z_{abs}=2.34$ (square, Cooke+2011a). The color shaded areas show
the results of the model for the three different populations of DLAs 
predicted to exist at $z=2.34$ (see Fig.~3).}
\end{centering}
\end{figure}
%%%%%%%%%%%%%%%%%%%%%%%%%%%%%%%%%%%%%%%%%%%%%%%%%%%%%%%%%%%%%%%%%%%%%%%%%%%%%%%

Finally, [Fe/H]$ > - 2$ DLAs are $T_{vir} > 10^4$~K halos forming
through merging of star-rich progenitors either before (Zone 4) or after 
(Zone 1) the end of reionization. Their systematically higher $Z$  
is a consequence of self-enrichment by a substantial ($M_* \approx 10^6-10^{9.5} \Msun$) 
population of long-living stars. Similarly, the high [C/Fe] ratios reflect 
the strong contribution by AGB stars. The gas mass in these 
DLAs is $M_g = 10^7-10^{9.5} \Msun$, thus resulting in higher $\nhi$ values, 
despite of the lower formation $z$ and larger $M_h\approx 10^{8-11} \Msun$ (eq.~1). 
Note however that the $\nhi$ derived for these DLAs must be interpreted as an upper limit. 
In fact, $M_h > M_{sf}$ objects (Fig.~3) are actively star-forming 
at $z=2.34$, with $\dot{M}\approx (0.1-10)\Msun$yr$^{-1}$, and 
hence part of their gas is presumably ionized. Moreover, the 
[Fe/H] value ([C/Fe]) has to be interpreted as a lower (upper) limit, 
since the contribution by SN type Ia may be significant in star-rich DLAs 
(Calura, Matteucci \& Vladilo 2002). The same interpretation has 
to be applied to [Fe/H] measurements, as Fe is depleted onto dust grains 
(Noterdaeme et al. 2008). Barred these limitations, included 
the unknown number of surviving satellites, we note
that the apparent concentration of [Fe/H]$>-2$ DLAs in a small region of the
$N_{HI}-Fe$ plane (Fig.~4) implies that these absorbers are only partially
representative of the progenitors of MW-like systems (Pontzen et~al. 2008). 
%%%%%%%%%%%%%%%%%%%%%%%%%%%%%%%%%%%%%%%%%%%%%%%%%%%%%%%%%%%%%%%%%%%%%%%%%%%%%%
\section{Discussion} 
%%%%%%%%%%%%%%%%%%%%%%%%%%%%%%%%%%%%%%%%%%%%%%%%%%%%%%%%%%%%%%%%%%%%%%%%%%%%%%
The recently discovered $z_{abs}\approx 2.34$ C-enhanced DLA pertains to a new class of systems, dubbed PopIII DLAs, hosting the first stars along with 
second generations of long-living stars. These systems are associated to 
H$_2$-cooling minihalos that virialize at $z > 8$ in neutral and primordial 
regions of the MW environment, and only experience a short period of SF before 
the increasing LW background burns out their H$_2$. Once ''sterilized'' these 
minihalos passively evolve as an inert gas reservoir, $M_g\approx 10^{5.5}-
10^{6.3}\Msun$, retaining the imprint of the stellar generations they 
hosted. Their gas temperature, $T_g$, is regulated by the balance between 
molecular/metal radiative cooling and photo-heating by the external ionizing 
radiation (Black 1981):
\begin{equation} 
n_{HI}\;\Lambda(T_g,Z) = \langle \epsilon \rangle K_{ph}
\end{equation}
where $n_{HI}$ is the gas density, $\Lambda(T_g,Z)$ the gas cooling rate 
(Maio et~al. 2007), $\langle \epsilon \rangle \approx 20$~eV the 
mean UV background photon energy, and $K_{ph}$ the optically-thick 
\HI photo-ionization rate (Abel \& Mo 1998). $K_{ph}\approx 4.1 
\times 10^{-15} J_{21} (N_{HI}/10^{19}$cm$^{-2})^{-\beta}$s$^{-1}$ with $\beta 
= 1.6$ and $J_{21}=1$. We find that [Fe/H]$ > - 6$ minihalos have $T_g \approx 
(40-1100) {\rm K} < T_{vir}$, implying that these systems can likely {\it survive} 
photo-evaporation thanks to self-shielding and metal cooling. Their gas 
temperature increases with decreasing metallicity, in agreement with the 
results by Kanekar et~al. (2009). The C-enhanced DLA has $T_g \approx 70$~K 
while in [Fe/H]$\approx -5$ ($Z\approx 10^{-4} \Zsun$) systems $T_g\approx 1000$~K. 

The peculiar abundance pattern observed in the 
$z_{abs}\approx 2.34$ DLA does not result from $Z = 0$ faint SNe, but rather 
from the enrichment by low-metallicity SNII and AGB stars, which may start 
to form as soon as $Z > \Zcr = 10^{-5\pm 1} \Zsun$. While SNII 
nucleosynthetic products are mostly lost in winds, AGB metals are retained 
in the ISM, causing a dramatic increase of [C/Fe]. The chemical 
evolution of [Fe/H]$> -5$ DLAs is {\it independent} on the assumed yields 
and IMF of PopIII stars, confirming the role of ordinary PopII stars in 
driving the enrichment of VMP systems (SSF07), recently emphasized by the
detection of a $Z \leq 5\times 10^{-5} \Zsun$ star with a "normal" chemical 
abundance pattern (Caffau et~al. 2011). If $Z_{cr} < 10^{-4}\Zsun$, as 
suggested by the existence of such star, the PopIII-PopII transition would 
be even quicker. The mass of relic stars in PopIII DLAs is found to be 
$M_*\approx 10^{2-4} \Msun$, implying that they are the gas-rich counterpart 
of the faintest dSphs. 

As stated, the C-enhanced DLA is a minihalo. However, we cannot exclude a different interpretation, in which such absorber might be a newly 
formed halo virializing at $z\approx 2.3$ from a rare, metal-free region of 
the Inter Galactic Medium, and actively forming PopIII stars.  
Since the total mass of the star-forming DLA is $M_h > 10^{9.5}\Msun
\approx M_{sf}(z=2.3)$, strong feedback is required to substantially remove 
the initial gas mass and match the observed $N_{\rm HI}$. By determining the 
dark matter content and SF rate of the C-enhanced DLA, it would be possible 
to disentangle these two pictures. 
We finally note that our simple semi-analytical model, which holds 
similarities with that proposed by Abel \& Mo (1998) for Lyman Limit Systems,
prevent us from making specific predictions on the number of DLAs at 
$z=2.34$. However, the relative contribution of C-enhanced DLAs to 
the total population is expected to be extremely low, $\approx 0.01\%$, thus
explaining the rarity of these systems.
%%%%%%%%%%%%%%%%%%%%%%%%%%%%%%%%%%%%%%%%%%%%%%%%%%%%%%%%%%%%%%%%%%%%%%%%%%%%%%%%
\section*{Acknowledgements}
We thank R. Cooke, P. Molaro, P. Petitjean, X. Prochaska \& R. Schneider 
for enlightening comments on the draft version of the paper. S.S. acknowledges 
a NOVA fellowship granted by the Netherlands Research School for Astronomy.
%%%%%%%%%%%%%%%%%%%%%%%%%%%%%%%%%%%%%%%%%%%%%%%%%%%%%%%%%%%%%%%%%%%%%%%%%%%%%%%%
 
\label{lastpage} 
\end{document}